\documentclass{article}
\usepackage{spconf,amsmath,epsfig}
\usepackage[numbers]{natbib}
\usepackage{caption}
\usepackage{subcaption}
\usepackage{color}

\usepackage[absolute,showboxes]{textpos}

\TPMargin{5pt}

\newcommand{\copyrightstatement}{
	\begin{textblock}{0.84}(0.08,0.93)    
		\noindent
		\scriptsize
		\copyright 2023 IEEE. Published in the IEEE 2023 International Geoscience \& Remote Sensing Symposium (IGARSS 2023), scheduled for July 16 - 21, 2023 in Pasadena, CA, USA. Personal use of this material is permitted. However, permission to reprint/republish this material for advertising or promotional purposes or for creating new collective works for resale or redistribution to servers or lists, or to reuse any copyrighted component of this work in other works, must be obtained from the IEEE. Contact: Manager, Copyrights and Permissions / IEEE Service Center / 445 Hoes Lane / P.O. Box 1331 / Piscataway, NJ 08855-1331, USA. Telephone: + Intl. 908-562-3966.
	\end{textblock}
}

\title{Reducing uncertainties of a chained hydrologic-hydraulic models to improve flood forecasting using multi-source Earth Observation data}
\name{T.~H.~Nguyen$^{1,2}$, S.~Ricci$^{1,2}$, A.~Piacentini$^{1}$, Q.~Bonassies$^{1,2}$, R.~Rodriquez~Suquet$^{3}$, S.~Pe\~{n}a~Luque$^{3}$, K.~Marlis$^{4}$ and C.~H.~David$^{4}$\thanks{Thanks to Space for Climate Observatory for funding.}}
\address{\small $^{1}$Centre Européen de Recherche et de Formation Avancée en Calcul Scientifique (CERFACS), 31057 Toulouse Cedex 1, France\\
    \small $^{2}$CNRS, CECI/CERFACS, UMR 5318, 31057 Toulouse Cedex 1, France\\
	\small $^{3}$Centre National d'Études Spatiales (CNES), 31401 Toulouse Cedex 9, France\\
    \small $^{4}$Jet Propulsion Laboratory, California Institute of Technology, Pasadena, CA, USA
 }

\begin{document}
\copyrightstatement

%
\maketitle
\begin{abstract}
The challenges in operational flood forecasting lie in producing reliable forecasts given constrained computational resources and within processing times that are compatible with near-real-time forecasting. Flood hydrodynamic models exploit observed data from gauge networks, e.g. water surface elevation (WSE) and/or discharge that describe the forcing time-series at the upstream and lateral boundary conditions of the model. A chained hydrologic-hydraulic model is thus interesting to allow extended lead time forecasts and overcome the limits of forecast when using only observed gauge measurements.
This research work focuses on comprehensively reducing the uncertainties in the model parameters, hydraulic state and especially the forcing data in order to improve the overall flood reanalysis and forecast performance. It aims at assimilating two main complementary EO data sources, namely in-situ WSE and SAR-derived flood extent observations. 
\end{abstract}

\section{Introduction}
Early warning and prediction of flood events have become all the more essential as the occurrence and intensity of flooding have increased in recent decades. 
In this regard, the Flood Detection, Alert and rapid Mapping \& Digital Twin project (labeled as FloodDAM-DT) has been supported by the Space for Climate Observatory, and partners together with the NASA ESTO/AIST Integrated Digital Earth Analysis System (IDEAS) project \cite{huang2022earth}. It pushes the developments of pre-operational tools to enable quick responses in various flood-prone areas while improving the description, reactivity and predictive capability \cite{kettig}, as well as combining numerical models and relevant observations to provide reliable predictions and projecting what-if scenarios and their impacts.
Relevant comprehension and assessment of flood hazards has thus become a pressing necessity, and the need for more reliable forecasts at extended lead times becomes all the more urgent. In this context, coupled hydrologic-hydraulic models are at the core of flood forecasting and flood risk assessments. 

In recent years, flood simulation and forecast capabilities have been greatly improved thanks to advances in data assimilation (DA) algorithms and strategies.
A classical DA approach stands in the assimilation of water surface elevation data, either from in-situ gauge measurements, from altimetry satellites, or retrieved from SAR or optical images using flood edge locations combined with complementary DEM data. 
DA combines in-situ gauge measurements and/or remote-sensing data with numerical models to correct the hydraulic states and reduce the uncertainties in the model state, parameters and forcings. The boundary conditions (BC), e.g. upstream inflow, consist in a time-series conventionally retrieved from gauge stations when available, thus limiting the forecast capability due to the transfer time of the river network. In order to reach longer lead times, forecasted inflow shall be provided by large-scale hydrologic models. For this matter, this paper focuses on a chained hydrologic-hydraulic model as it allows for extended lead time forecasts. 
Hydrologic model outputs can also provide inputs for local-scale models in poorly-gauged catchments.


\cite{nguyen2016high} proposes a coupled hydrologic-hydraulic model called HiResFlood-UCI for flash flood modeling. It combines the hydrologic model HL-RDHM developed by US National Weather Service with a 2D hydraulic model BreZo. The HL-RDHM was used as a rainfall-runoff generator and then its routing scheme was replaced by the BreZo model in order to predict flood depths and velocities at a local scale.
\cite{grimaldi2019challenges} focused on quantifying the propagation of errors when coupling a hydrological model HBV and a hydraulic model LISFLOOD‐FP for a seven-year flood analysis on a large basin featuring morphological and hydrological variability.
They emphasized the sensitivity of inundation volume and extent prediction under the effect of uncertainties in flood peak values. 
\cite{hostache2018near} presents a data assimilation of SAR-derived flood probability maps to reduce the uncertainty of a flood forecasting model cascade composed of the conceptual hydrological model SUPERFLEX loosely coupled with the grid-based hydraulic model LISFLOOD-FP. Such models were shown   computationally efficient and able to capture the streamflow and floodplain dynamics, making them suited for near-real-time flood forecasting. A similar coupling strategy is used by \cite{DiMauro2021} while focusing on mitigating the degeneracy of the Particle Filters.
%

\begin{figure}[t]
    \centering
    \includegraphics[width=0.5\textwidth]{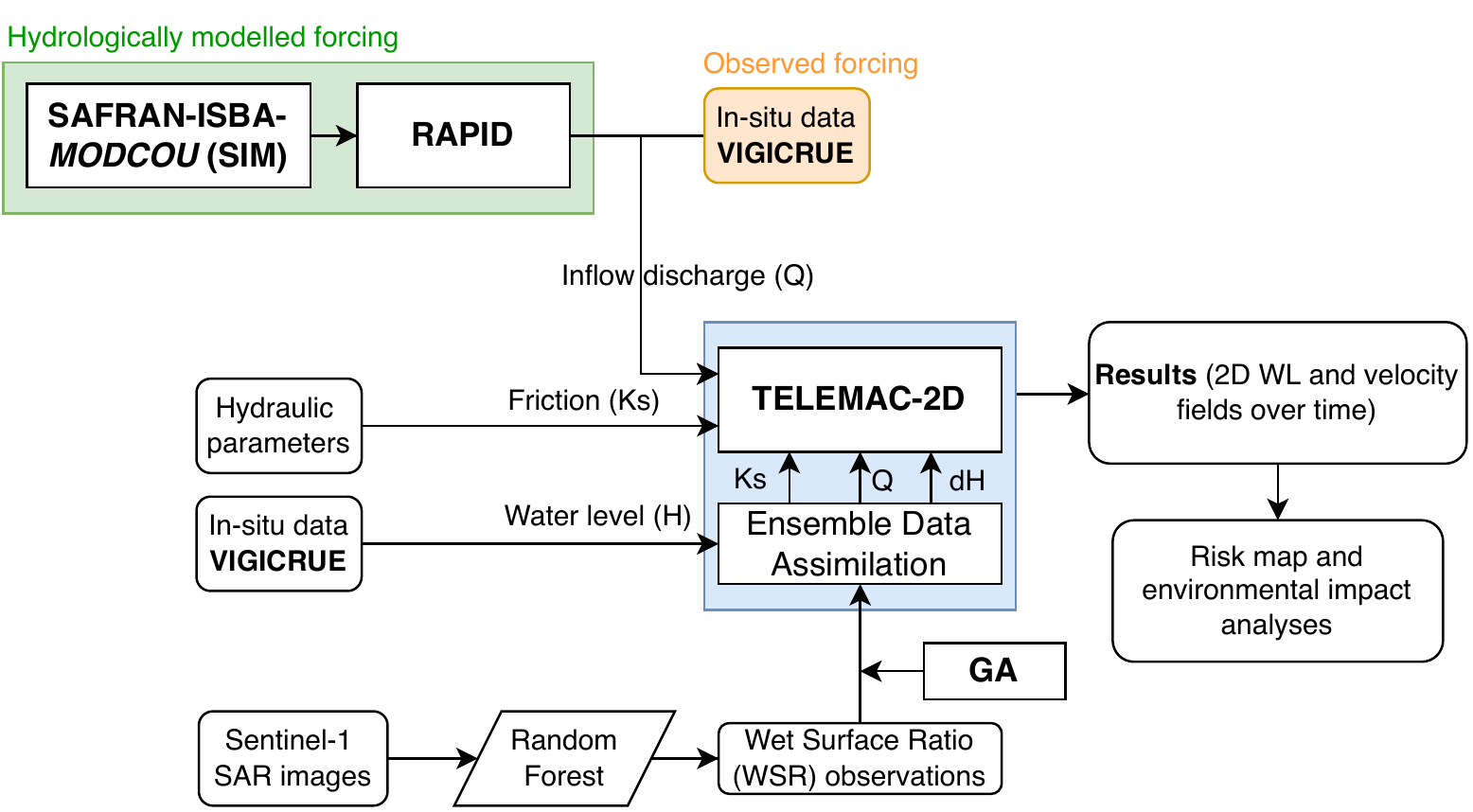}
    \caption{Overview workflow.}
    \label{fig:ExpSetWorkflow}
\end{figure}

\section{Data and Models}
\label{sec:data}
The study area is the Garonne Marmandaise catchment (southwest of France) which extends over a 50-km reach of the Garonne River between Tonneins and La Réole. The available observations are composed of the WSE time-series at the three observing stations, as well as the wet surface ratios (WSR) computed over five floodplain subdomains based on the flood extent maps derived from Sentinel-1 SAR images.

Routing Application for Parallel Computation of Discharge, abbreviated as RAPID, is a numerical model that simulates the propagation of water flow waves in networks of rivers made of tens to hundreds of thousands of river reaches. The river routing approach in RAPID is based on the traditional Muskingum routing method adapted to a matrix-based formulation to compute discharge simultaneously through a river network \cite{david2011river}. 
Given the inflow of water from land and aquifers---as commonly computed by land surface models (LSMs)---RAPID can simulate river discharge at the outlet of each river reach of large river networks. 
When applied over France, RAPID is forced by LSM productions from the SAFRAN-ISBA hydrometeoreological model \cite{habets2008safran}, and it replaces the MODCOU hydrogeological model in the SAFRAN-ISBA-MODCOU (SIM) suite. 
In the present work, the BC for a high-fidelity and local-scale hydrodynamic model with TELEMAC-2D (T2D) are provided by the RAPID model.

T2D is a parallel numerical solver of the shallow water equations---derived from the free surface Navier-Stokes equations---with an explicit first-order time integration scheme, a finite element scheme, and an iterative conjugate gradient method \cite{hervouet2007hydrodynamics}.
As a hydraulic model, it takes into account the amount of water entering the river system to compute the WSE and velocity in the river network, and when the storage capacity of the river is exceeded, in the flood plain. These models are needed in order to predict water depth, velocity and flood extent that are fundamental to predict, prevent and/or mitigate devastating consequences of floods.

\section{Method}
\label{sec:method}

\begin{figure}[!]
    \centering
    \includegraphics[width=0.5\textwidth]{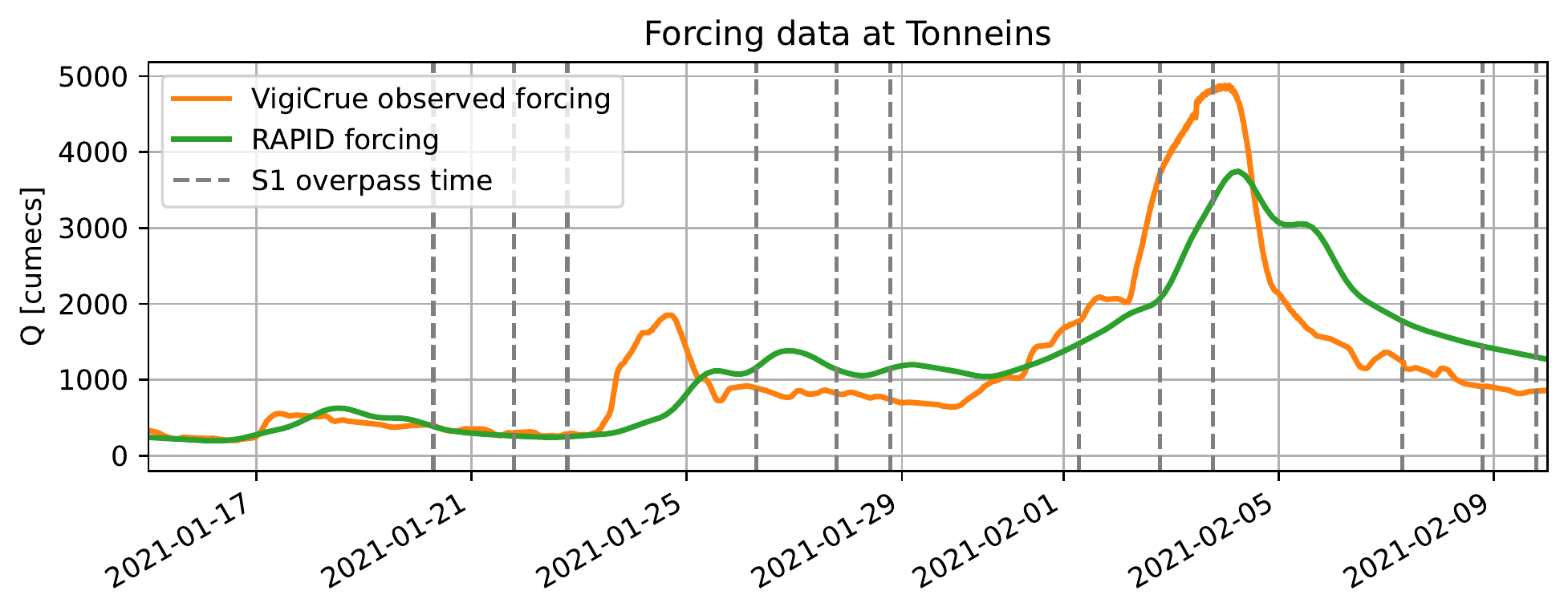}
    \caption{Observed (orange) and RAPID (green) discharge time-series.}
    \label{fig:inflow}
\end{figure}

An overview workflow of this research work is illustrated in Fig.~\ref{fig:ExpSetWorkflow}, in which the hydrological forcing data is either from observed streamflows (from an in-situ station, orange box) or provided by hydrologic model (green box). They are depicted by the same color in Fig.~\ref{fig:inflow}, as well as the Sentinel-1 overpass time by vertical dashed lines during this event.
It should be noted that the hydrologic model SAFRAN-ISBA-RAPID generally yields better performances for large basins and that a sub-optimal performance could be expected for the medium-sized Garonne catchment, especially for high flow regimes, as evidenced by Fig.~\ref{fig:inflow}. This could stem from the uncertainties in its land surface inputs, parameters and catchment description. As a result, the water discharge tends to be underestimated by the hydrologic model for major flood events.
Subsequently, major sources of uncertainties in the chained hydrologic-hydraulic model thus lie in the RAPID-simulated discharges as well as in the T2D parameters, e.g. friction, floodplain topography and river bed geometry. Fig.  

In this work, a sequential ensemble DA is applied on top of the hydraulic T2D model to deal with the uncertainties in friction, forcing and hydraulic state with a time-varying correction with a dual state-parameter EnKF analysis. 
It involves the assimilation of 2D flood observations derived from remote-sensing images (typically Sentinel-1 SAR images, and Sentinel-2 optical images if relevant) over the Garonne Marmandaise catchment. A Random Forest algorithm is applied to derive flood extent maps from the available images acquired during overflowing events \cite{NguyenTGRS2022}. The resulting binary wet/dry maps are then expressed in terms of so-called wet surface ratios (WSR) over selected subdomains of the floodplain. This ratio is assimilated jointly with in-situ WSE observations to improve the flow dynamics within the floodplain, as proposed in previous work \cite{nguyenagu2022}. The performed DA is also improved by a Gaussian anamorphosis (GA) transformation \cite{nguyenagu2022,nguyen2023gaussianarxiv} to deal with non-Gaussian errors in the WSR observations that could hinder the optimality of the EnKF.

\begin{figure}[t]
    \centering
    \begin{subfigure}{0.44\textwidth}
    \centering
    \includegraphics[trim=0 1.8cm 0 0,clip,width=\textwidth]{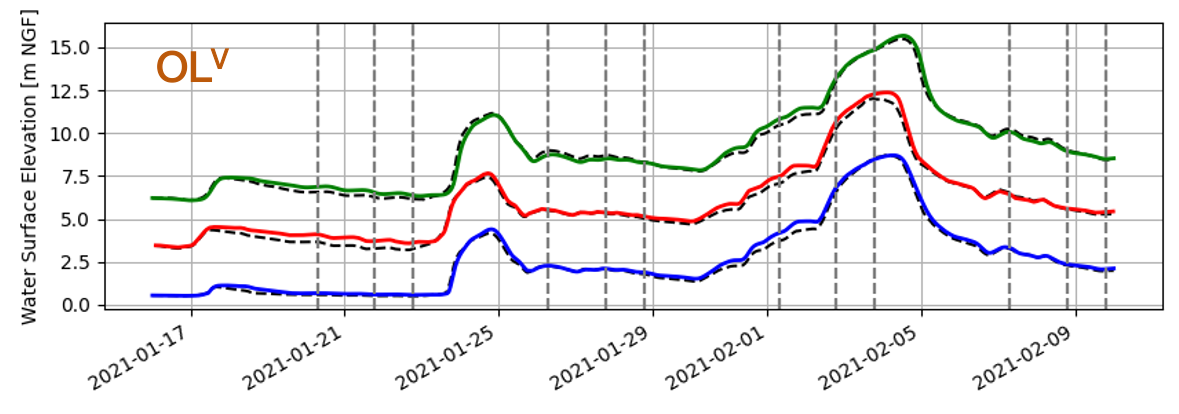}
    \includegraphics[width=\textwidth]{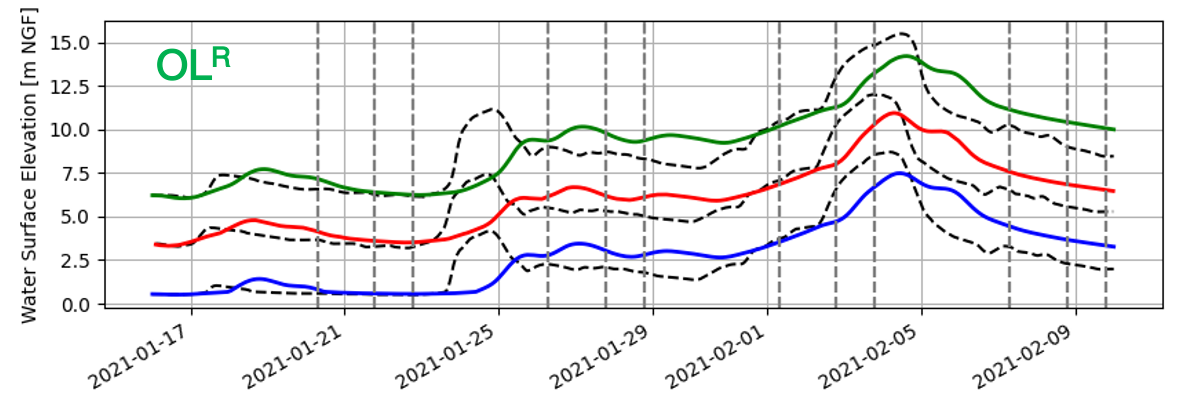}
    \caption{OL using observed (top) and RAPID discharge (bottom).}\label{fig:WSE_FR}
    \end{subfigure}\hfill
    \begin{subfigure}{0.44\textwidth}
    \centering
    \includegraphics[trim=0 1.8cm 0 0,clip,width=\textwidth]{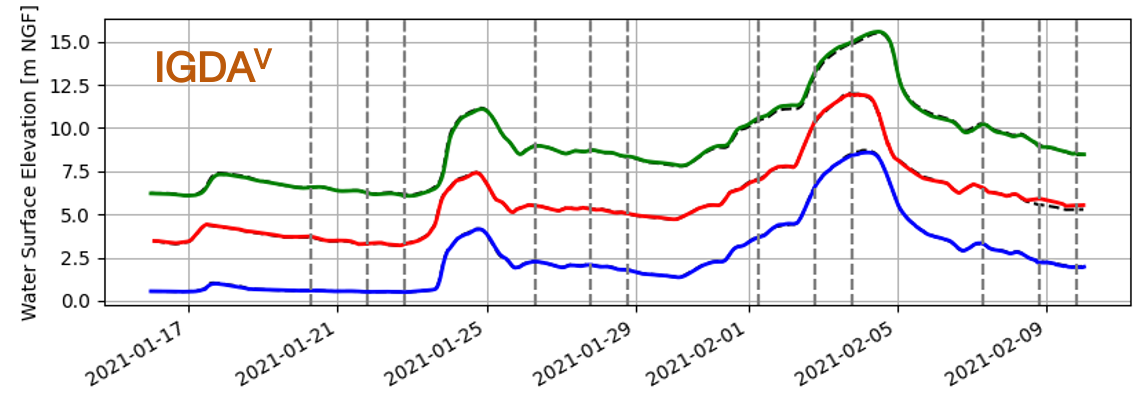}
    \includegraphics[width=\textwidth]{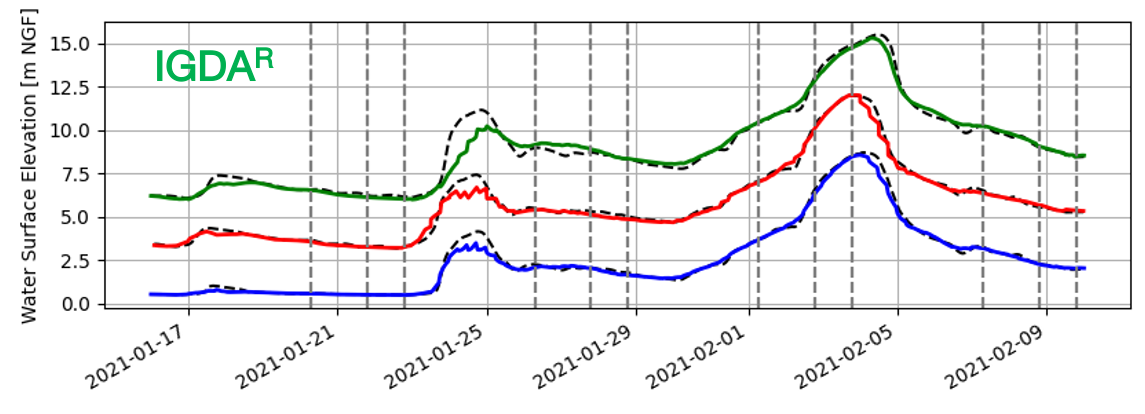}
    \caption{IGDA using observed (top) and RAPID discharge (bottom).}\label{fig:WSE_IGDA}
    \end{subfigure}
    \caption{Simulated WSE by OL and IGDA experiments using observed and hydrologically-modeled discharge at in-situ observing stations Tonneins (red), Marmande (blue) and La Réole (green) with respective observed WSE (dashed lines).}
    \label{fig:WSE}
\end{figure}

Full details on the performed cycled EnKF can be found in \cite{NguyenTGRS2022,nguyenagu2022}. The error in the RAPID forcing is taken into account through a multiplicative factor $\mu$ on the time-dependent discharge time-series. Also, the error in the hydraulic state that is due to the lack of evapotranspiration, ground infiltration and rainfall processes is taken into account as a state correction  implemented over sub-domains of the floodplain, denoted by $\delta H$. The observation operator associated with the WSR and in-situ observations, the GA-assisted EnKF algorithm with a dual state-parameter sequential correction are devised and assessed using metrics that are formulated in 1D at the observing stations in the river bed and 2D over the floodplain.

The proposed method is validated in a pseudo-twin experiment setting, in which the studied flood event was created by a set of parameters taken from the real 2003 flood event, resulting in a deterministic reference simulation. In addition, we shifted the time frame of this setting to correspond to the 2021 flood event occurred over the same catchment, in order to re-use Sentinel-1 time frame (i.e. only the time, not the observations). As such, the observation data was synthesized from the reference simulation. This stands in the extraction of the true simulated WSE at all observation times and locations, first to generate synthetical in-situ observations, and second to extract the wet/dry pixels to compute the WSR. The forcing data, however, is from the real flood event, either observed by the in-situ station at Tonneins (the upstream point of the T2D model) or modeled by the hydrologic RAPID model for the Garonne reach close to Tonneins.

\begin{figure}[t]
    \centering\includegraphics[width=0.85\linewidth]{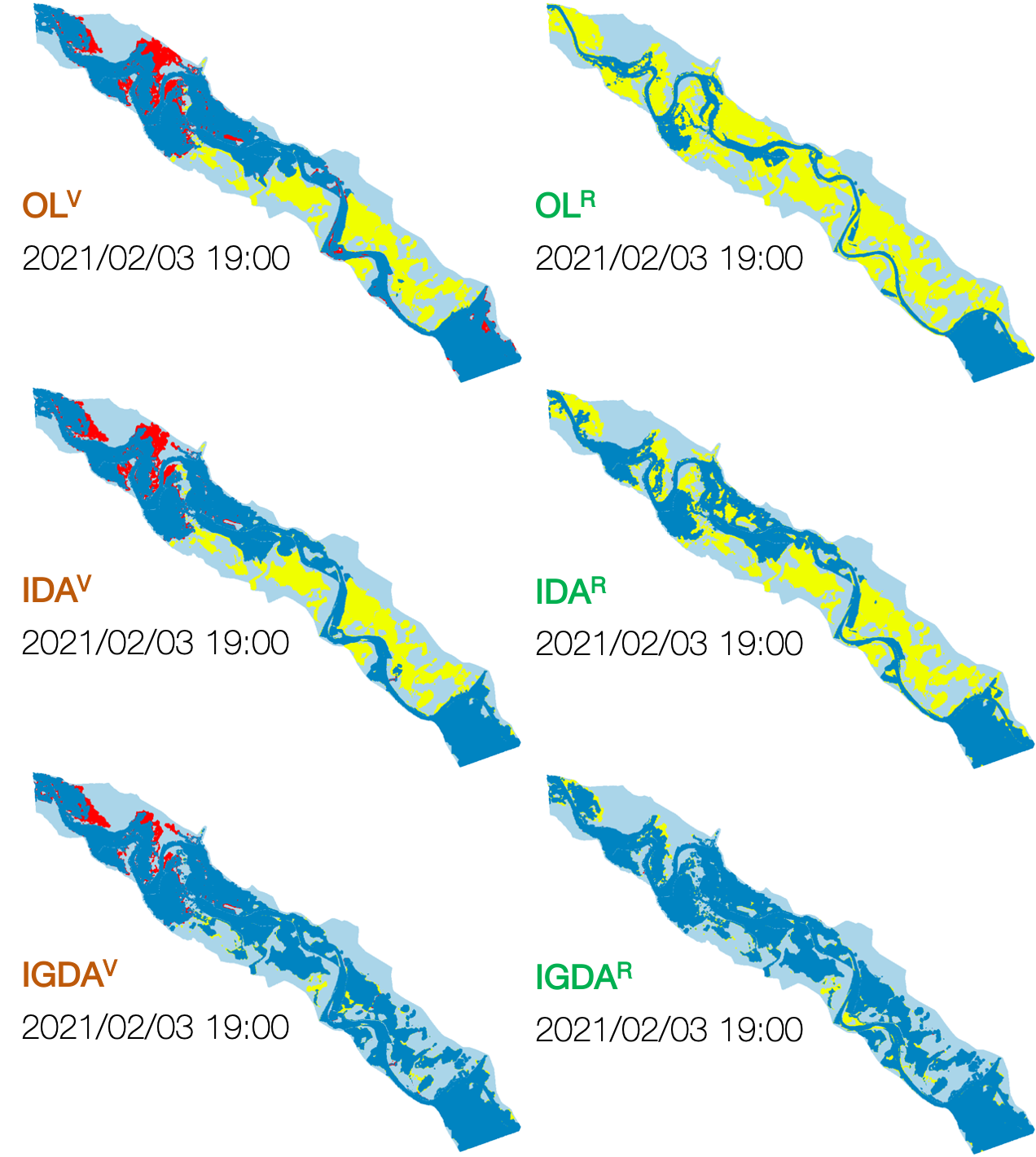}
    \includegraphics[width=0.5\linewidth]{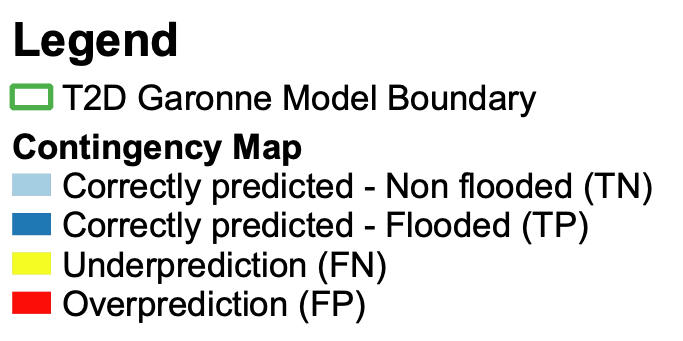}
    \caption{Contingency maps between simulated flood extent and observed flood extent derived from S1 image.}\label{fig:2D-assessment}
\end{figure}

\section{Result and Discussion}
\label{sec:result}

\begin{table*}[t]
        \centering
        \small
        \caption{1D RMSE with respect to WSE measured at in-situ stations, and 2D CSI with respect to observed flood extent maps.}\label{tab:result_2003}
        \begin{tabular}{c|cc|ccc|ccc}
            \hline
            & Assimilated Obs. & Control Vector & \multicolumn{3}{c|}{{Root-mean-square Error [m]}} & \multicolumn{3}{c}{{Critical Success Index [\%]}}\\
            \hline
            \textbf{Exp.} & & & Tonneins & Marmande & La Réole & 2021/02/02 & 2021/02/03 & 2021/02/07 \\
            \hline
            {OL$^\mathrm{V}$} & - & - & 0.359 & 0.193 & 0.225 & 49.65 & 67.90 & 74.53 \\
            \hline
            {IDA$^\mathrm{V}$} & in-situ WSE & friction + $\mu$ & 0.053 & 0.036 & 0.080 & 48.67 & 68.30 & 76.10 \\
            \hline
            {IGDA$^\mathrm{V}$} & in-situ WSE + WSR & friction + $\mu$ + $\delta H$ & 0.059 & 0.035 & 0.087 & 95.41 & 92.32 & 88.28 \\
            \hline
            \multicolumn{9}{c}{}\\
            \hline
            {OL$^\mathrm{R}$} & - & - & 1.550 & 1.254 & 1.370 & 46.06 & 36.63 & 63.24 \\
            \hline
            {IDA$^\mathrm{R}$} & in-situ WSE & friction + $\mu$ & 0.467 & 0.292 & 0.635 & 48.77 & 57.90 & 77.63 \\
            \hline
            {IGDA$^\mathrm{R}$} & in-situ WSE + WSR & friction + $\mu$ + $\delta H$ & 0.326 & 0.229 & 0.440 & 95.76 & 94.34 & 88.38 \\
            \hline
        \end{tabular}
    \end{table*}
    
Three experiments are carried out using either observed or RAPID-simulated forcings: an open-loop (OL) without assimilation and two DA experiments, namely IDA that assimilates only in-situ WSE measurements and IGDA that assimilates both in-situ WSE and remote-sensing WSR observations. 
In the following, superscript attached to the experiment name indicates the used forcing data, for instance, IDA$^\mathrm{V}$ stands for the IDA experiment forced by observed VigiCrue discharge, whereas IGDA$^\mathrm{R}$ represents the IGDA experiment forced by RAPID simulated hydrograph.
Quantitative results are presented in Table~\ref{tab:result_2003}, with 1D metrics computed with respect to WSE at observing stations located along the river and 2D metrics computed with respect to reference/observed flood extents, namely Critical Success Index (CSI). 
Fig. \ref{fig:WSE} depicts the simulated WSE (continuous lines) and observed WSE (dashed lines) at the observing stations, when using the two different inflow discharge time-series, namely Fig.~\ref{fig:WSE_FR} and Fig.~\ref{fig:WSE_IGDA}, respectively, for OL  and IGDA experiments with both discharges. For both observed and RAPID-simulated upstream forcings, the joint assimilation of in-situ WSE and RS-derived WSR observations in IGDA leads to a significant improvement of the 1D and 2D metrics with reduced RMSE and increased CSI. 
Fig. \ref{fig:2D-assessment} reveals the contingency maps  computed with respect to the reference flood extent maps near the flood peak on 2021/02/03. It is shown that IGDA with Gaussian anamorphosis further improve the results, especially in terms of representation of the flood dynamics.

Finally, this work demonstrates that while the forcing provided by a large-scale hydrologic model can be seamlessly used as input to a local hydraulic model, and DA allows for an efficient reduction of the uncertainty in the hydrology products. These conclusions advocate a multi-source strategy for the assimilation algorithm implemented on top of a chained hydrologic-hydraulic model, making the most of heterogeneous data with different spatial and temporal resolution and distribution.

\footnotesize

\bibliographystyle{ieeetr}
\bibliography{refs}

\end{document}